\documentstyle[12pt]{article}
\begin{document}
{\pagestyle{empty}
\vskip 2.5cm
~~\\

{\renewcommand{\thefootnote}{\fnsymbol{footnote}}
\centerline{\large \bf Simulated Annealing with Tsallis Weights}
\centerline{\large \bf A Numerical Comparison}
}
\vskip 3.0cm

\centerline{Ulrich H.E.~Hansmann\footnote{\ \ e-mail: hansmann@ims.ac.jp}}

\centerline{\it Department of Theoretical Studies}
\centerline{\it Institute for Molecular Science}
\centerline{\it Okazaki, Aichi 444, Japan}

\medbreak
\vskip 3.5cm
 
\centerline{\bf ABSTRACT}
\vskip 0.3cm
We discuss the use of Tsallis generalized mechanics in simulated annealing
algorithms. For a small peptide it is shown that older implementations are
not more effective than regular simulated annealing in finding
ground state configurations. We propose a new implementation which leads
to an improvement over  regular simulated annealing.
\vskip 3.5cm

\vfill
\newpage}
\baselineskip=0.8cm
%%%%%%%%%%%%%%%%%%%%%%% Section 1 %%%%%%%%%%%%%%%%%%%%%%%%
Simulated annealing \cite{SA} has become an often used tool to tackle 
hard 
optimization problems in various fields of science. Its underlying idea of 
modeling the crystal grow process in nature is  easy to
understand and simple to implement. Any Monte Carlo or molecular dynamics
technique can be converted into a simulated annealing algorithm by 
allowing the
temperature to decrease gradually. However, the method is not without
problems. The performance of simulated annealing depends crucially on the
annealing schedule. It could be shown that convergence to the global
minimum can be secured for a logarithmic annealing schedule \cite{GG},
but this is of little use in  applications of the method. 
Constraints in available computer  time enforce the choice of faster 
annealing schedules where
success is no longer guaranteed. 
Similar to the growing of real crystals which hardly ever can be archived
by a simple cooling process, elaborated and often system specific 
annealing schedules are frequently
necessary to obtain the global minimum in the CPU time available. 
Hence, ever since the
seminal article by Kirkpatrick {\it et al.} \cite{SA}, attempts were made,
to improve the performance of simulated annealing in practical 
applications, see for instance Ref.~\cite{FSA}. Two of the more recent 
attempts \cite{penna,straub} are inspired by Tsallis generalized mechanics
\cite{tsa,TSA}. We compare the performance of both implementations with regular 
simulated annealing by taking an energy function for the protein folding
problem as an example. We describe then a new application of Tsallis weights to
simulated annealing which leads for our model to an improvement over
regular simulated annealing.

In the Tsallis formalism \cite{tsa}, a generalized mechanics is constructed by
maximizing a generalized entropy
\begin{equation}
S = -k \sum _i p_i \ln p_i
\end{equation}
with the  constraints 
\begin{equation}
\sum_i p_i = 1 \quad \quad \sum_i p_i^q E_i = {\rm const}~.
\end{equation}
Here, $q$ is a real number. 
A generalized probability distribution
\begin{equation}
p_i \propto \left[ 1- (1-q)\beta E_i \right]^{\displaystyle \frac{1}{1-q}}
\label{Tdis}
\end{equation}
follows and the average of an observable ${\cal{O}}$ can be defined by 
\begin{equation}
<{\cal{O}}> = \sum_i p_i^q {\cal{O}}_i~.
\end{equation}
It is evident that for $q \rightarrow 1$ the generalized distribution of 
Eq.~\ref{Tdis} 
tends toward the  Gibbs-Boltzmann distribution and therefore regular 
statistical mechanics is recovered in this limit.

The important feature of Tsallis generalized statistic for optimization
problems is that  the probability of states does no longer decrease 
exponentially with energy but according to a  power law where 
 the exponent is determined by the  free parameter $q$ (see Eq.~\ref{Tdis}).
This observation inspired a generalized simulated
annealing algorithm \cite{penna} where the acceptance probability
\begin{equation}
p(\Delta E) = \min \left[ 1, [1-(1-q)\beta \Delta E ]^{\frac{1}{1-q}} \right]
\label{eqpen}
\end{equation}
was introduced. 
Here, $\Delta E $ is the change in  energy. Again, for $q \rightarrow 1$,
the acceptance probability of canonical simulated annealing is recovered.
The algorithm was employed to find close to
optimal solutions to the traveling salesman problem and it was claimed
that it is faster than classical simulated annealing and has optimal
performance for large negative parameters $q$ \cite{penna}.  
However, the above algorithm 
does not obey detailed balance and therefore convergence to an equilibrium 
distribution  is in general not guaranteed. For this reason, it 
was recently proposed to utilizes the acceptance probability \cite{straub} 
\begin{equation}
p(E_{old} \rightarrow E_{new}) = \min \left[ 1,\left(
\frac{1-[1-q(T)]\beta E_{new}}{1-[1-q(T)]\beta E_{old}} 
                  \right)^{\displaystyle \frac{q(T)}{1-q(T)}}
                                      \right]
\label{STR}
\end{equation}
with $\lim_{T\rightarrow 0} q(T)=1$.  
Detailed balance is obtained by the algorithm and  convergence to the 
generalized  distribution of Eq.~\ref{Tdis} is guaranteed   
 for each temperature $T$ and parameter $q$.  Since $q \rightarrow 1$ as
$T \rightarrow 0$, this generalized simulated annealing algorithm
tends in the same way as regular simulated annealing to a steepest descent
at low temperature. 
For  Tsallis
parameters $q(T) > 1$ the distribution of Eq.~\ref{Tdis} 
 has a tail to higher energies and  
the probability to cross energy barriers and to escape local minima 
is therefore increased by the above weights. 
The new  algorithm 
was applied in Ref.~\cite{straub} 
to the conformational optimization of the 48 atom
tetraalanine peptide using molecular dynamics and hybrid MD-MC
methods in the CHARMM force field \cite{CHARMM}. Both temperature and
the Tsallis parameter $q$ were exponentially decreased, with 
a  start value of $q=2$ for the Tsallis parameter. Results better than
conventional simulated annealing were reported. 

However, a disadvantage of the above algorithm is that it  requires the 
careful tuning of additional free parameters. Not only a suitable  annealing
schedule in temperature $T$ has to be chosen, but also one in the
Tsallis parameter $q(T)$. Furthermore, the performance of this algorithm 
depends also  on the choice of the zero in
potential energy  
(which was ignored by the authors of Ref.~\cite{straub}).  
This is because the transition 
probabilities of Eq.~\ref{STR}  are not invariant under a shift in energy. 
For instance, even for $q > 1$
(the case $q< 1$ can lead to complex probabilities) the weights of
 Eq.~\ref{STR}  can become negative for negative values of the energies. 
 On the other hand,  acceptance of a configuration will no longer
depend on the energy of the configuration if all energies are shifted by a
large enough positive number.
Hence, while classical simulated annealing has already the
problem of finding an optimal annealing schedule for the temperature,
the above algorithm requires in addition a careful tuning of the
Tsallis parameter $q$ and of the energy scale, making its applicability
highly model dependent.

Here we show how the above problems can be alleviated.  Use of Tsallis 
weights  in the course of a simulated annealing simulation is motivated by the
fact that the resulting probability distribution has a tail to 
higher energies for $q > 1$, enhancing in this way the probability to
cross barriers and escape local minima. While negligible at high
temperatures this feature becomes important at low temperatures
where a canonical weight would make it difficult to escape local
minimas. We are therefore mainly interested in the use of of Tsallis
weights for {\it low} temperatures. It is obvious that the 
Tsallis distribution at low temperatures 
should not be dominated by the tail to higher energies, 
but still be centered around the energy where the canonical
distribution has its maximum. For otherwise, low energy (temperature) 
states will not be  sampled sufficiently.
To find the parameter $q > 1$ which yields to an optimal distribution 
let us first write the Tsallis weights as 
\begin{equation}
w(E) = \left[ 1- (1-q)\beta (E - E_{0}) \right]^{\displaystyle \frac{q}{1-q}}~,
\label{tsw2}
\end{equation}
where $E_{0}$ is the (in general unknown) ground state energy. This is
equivalent to chosing an energy scale where all energies are positive with
the zero for the ground state.  In this way we ensure that the weights 
are always positive. The Tsallis weights will be a good approximation 
 of the Boltzmann weights 
$ W_{B}(E) = \exp(-\beta (E-E_{0})) $
for \hbox{$(1 -q) \beta (E - E_{0}) << 1~$}.
 To ensure that simulations are able to escape from
energy local minima, the weights should start deviating from the
exponentially damped Boltzmann weights at energies near its mean value.
This is because at low temperatures there are only small fluctuations of
energy around its mean. We may thus set in Eq.~\ref{tsw2}:
\begin{equation}
-(1-q)\beta \left( <E>~-~E_0 \right) = \frac{1}{2}
\label{Tcon}
\end{equation}
The mean value of energy is given at low temperatures by the harmonic 
approximation:
\begin{equation}
<E>-E_{0} \approx \frac{n_F}{2} k T = \frac{n_F}{2\beta}~,
\end{equation}
where $n_F$ is the degree of freedom of our molecule. 
Hence, for low temperatures, Eq.~\ref{Tcon} can be written, as
\begin{equation}
-(1-q) \frac{n_F}{2} = \frac{1}{2}~,
\end{equation}
which leads to an optimal Tsallis parameter 
\begin{equation}
q = 1 + \frac{1}{n_F}~.
\label{qtsa}
\end{equation}

Hence, we  propose a generalized simulated annealing algorithm where
configurations are weighted with
\begin{equation}
w(E) = 
\left[ 1- \left( 1-q\right) \beta \left( E - E_{0}\right) 
    \right]^{\displaystyle \frac{q}{1-q}}~.
\label{wfinal}
\end{equation} 
The Tsallis parameter $q$ is  set to $q=1+1/n_F$.
 Our weights require knowledge of the 
ground state energy. However, in general $E_0$ is not known.
We therefore approximate $E_0$  in the course of a simulated
annealing simulation by $E_{0} \equiv E_{min} -c$ where  
$E_{min}$ is the
lowest energy ever encountered in the simulation and $c$ a small number.
$E_0$ is reset every time a new value for $E_{min}$
is found. Changing the value of $E_{0}$ is a
disturbance of the Markov chain and while we expect the disturbance to be 
small, we clearly cannot use our algorithm to calculate thermodynamic 
averages. However, due
to finite stepsize of the temperature annealing we can anyway not
assume convergence against an equilibrium distribution. As with regular
simulated annealing, our method is  valid only as a global optimization
method.

We have tested the various simulated annealing algorithms for the 
protein folding problem, a long-standing problem in
biophysics with rough energy landscape. Here, Met-enkephalin has become
a often used  model to examine new algorithms. 
Met-enkephalin has the amino acid sequence Tyr-Gly-Gly-Phe-Met.
The potential energy function $E_{tot}$ that we used is given 
by the sum of 
electrostatic term $E_C$, Lennard-Jones term $E_{LJ}$, and
hydrogen-bond term $E_{hb}$ for all pairs of atoms in the peptide
together with the torsion term $E_{tors}$ for all torsion angles.
The parameters for the energy function were adopted from
ECEPP/2.\cite{EC3}  Fixing the peptide bond angles $\omega$ to 
$180^{\circ}$ leaves us with 19 torsion angles as degree of freedom.
 The computer code KONF90 \cite{KONF}  was
used.  

As in earlier work on Met-enkephalin \cite{ho94_3} we performed for each 
of the different algorithms 
 20 runs of 50,000 sweeps. Each run started from completely random
configuration and  each angle is updated once in a sweep.  The temperature 
was lowered exponentially according to
\begin{equation}
T = T_{ST} \gamma^{i-1}
\end{equation}
for the $i$th sweep and 
\begin{equation}
\gamma = (T_{FI}/T_{ST})^{\frac{1}{49999}}~,
\end{equation}
where $T_{ST}$ is the start temperature and $T_{FI}$ is the final temperature.
One of the
quantities we monitored to evaluate the performance of the various
algorithms was the average $<E_{Low}>$ (taken over all 20 runs) of the 
lowest energies $E_{Low}$ 
obtained in each single run. The other quantity was the number $n_G$  
of ground-state configurations found in the 20 independent runs.
 In Ref.~\cite{EnkO}
it was shown that with the energy function KONF90, conformations
of energy less than
$-11.0$ kcal/mol have essentially the same three-dimensional
structure.  Hence, we consider  any conformation with
$E \le -11.0$ kcal/mol as the ground-state configuration. 
 
Let us present now our results. In Tab.~1 we show typical results for
the first generalized simulated annealing algorithm  which was proposed in
Ref.~\cite{penna} and 
uses the acceptance probability of Eq.~\ref{eqpen}. We chose as start
temperature $T_{ST}=1000~K$ and  tried two values for the final
temperature: $T_{F} =50~K$ and $T_F=1~K$. These temperatures were also
chosen by us in Ref.~\cite{ho94_3} from which we have also taken the 
results for regular simulated annealing (indicated  by $q=1$ in Tab.~1).
We did not  
observe an improvement over regular simulated annealing for any choice 
of the Tsallis parameter $q$ in this generalized simulated annealing 
algorithm. In the range $0\le q \le 1.25$ the performance 
is comparable with canonical simulated annealing. Both $n_G$ and $<E_{Low}>$
vary little in this range. The performance of the algorithm detoriates  
quickly outside of this range. This is especially true for large
negative values of $q$ which were presented as the optimal choice for $q$ in 
Ref.~\cite{penna}. The probability of finding ground state configurations
becomes small and  the average $<E_{Low}>$ of lowest energies found in each 
single run not only increases, but also the standard deviation of $<E_{Low}>$
indicating that the obtained low energies strongly depend on
the initial configuration  and the algorithm reduce in this case 
to a mere quenching. Hence, for our system, application of this
generalized simulated annealing algorithm seems to bring no improvement over
regular simulated annealing.

In Tab.~2 we show our  results for the second method, which uses the 
acceptance probability
of Eq.~\ref{STR} and  was proposed in Ref.~\cite{straub}.
Following the authors of Ref.~\cite{straub} we chose  the start value
$q=2$ and decreased both temperature and Tsallis parameter exponentially
such that for the final temperature $T_{FI}$ we have $q(T_{FI})=1$. Again we  
tried two values for the final temperature: $T_{F} =50~K$ and $T_F=1~K$. 
Since Tsallis distributions have a tail to high energies for $q >1$, we
expected that it would be possible to choose lower start temperatures
than for  regular simulated annealing.  Hence, we tried not only
$T_{ST}= 1000~K$, but also $T_{ST}=500~K$ and $300~K$.
 In each case we found a
poorer performance than   for regular simulated annealing. This is in
contradiction to the results of Ref.~\cite{straub}, where the authors found
a significant improvement over canonical simulated annealing. We remark
that for the calculation of weights (see Eq.~\ref{STR}) in the simulations, 
we shifted the energies  by the  ground-state
energy for Met-enkephalin (as known from previous work \cite{ho94_3}) 
to ensure positive weights, otherwise  even poorer
results were obtained. Hence, the poor performance of this algorithm 
in simulations of our peptide is
not only due to a poor choice of the energy scale, but also to an imperfect 
annealing schedule of the Tsallis parameter $q(T)$. 
We conclude that the performance of this algorithm is
highly model dependent and requires carefully tuning in the annealing
of $q$ and the choice of the energy scale. This is a severe limitation for 
practical applications.

Finally, Tab.~3  shows the results for our implementation of Tsallis weights
in simulated annealing algorithms  using the acceptance probability of
Eq.~\ref{wfinal}. The main difference to the  previous algorithm is 
that the 
Tsallis parameter $q$ is not free, but set to an optimal value of  
 $q_F = 1 + 1/n_F = 1+1/19 \approx 1.053$. In addition, 
 our procedure guarantees that the weights are always positive by 
self-tuning the estimate of the ground state energy $E_0$ in the course of
an annealing run.  $E_0$ is reset every time to  $E_0 = E_{min}- 1~kcal/mol$
when a new configuration with lower energy $E_{min}$  than  any previous
configuration is found. 
We found  for  both canonical and 
generalized simulated annealing an optimal performance for
 start temperature $T_{ST} =500~K$ and 
final temperature $T_{F}=50~K$. With this  temperature
annealing schedule the ground state configuration was found 8 out of 20 runs 
for regular simulated annealing and 12 out of 20 runs for generalized
simulated annealing. This is a modest improvement of the new algorithm
over the canonical simulated annealing. The improvement can also be seen in the
estimate for $<E_{Low}>$ which is $0.6$ kcal/mol lower for the new
algorithm and has a smaller standard deviation than regular simulated 
annealing. We
further notice that the new generalized ensemble algorithm allows 
to start the temperature annealing at lower temperatures. While
regular simulated annealing works best with start temperatures over
$500~K$, the performance of the new algorithm depends only little on the start
temperature and rather favors $T_{ST} \le 500~K$. This
follows from the form of the Tsallis distributions which have a tail to 
high energies for $q > 1$. Equilibrization at lower
temperatures is therefore enhanced.  
We remark that the improvement of the new method over regular 
simulated annealing still does not lead to 
the performance
reported for the  generalized ensembles algorithms in 
Refs.~\cite{ho94_3,HO96b}.  However, the new algorithm is much easier to
implement. Since the Tsallis parameter $q$ is  constant  in our algorithm,
 only minor modifications are required in existing simulated annealing
programs to accommodate the new technique.
Unlike in the algorithm of Ref.~\cite{straub} the improvement over regular
simulated annealing is gained without the need of determining optimal
annealing schedules for additional parameters.

Let us summarize our results. 
We have performed Monte Carlo simulations of Met-enkephalin using
Tsallis generalized Mechanics. We  discussed older proposals
for the use of Tsallis weights in simulated annealing algorithms
and showed how to overcome their shortcomings.  Our algorithm is
easy to implement in existing simulated annealing programs and does not
require tuning of annealing schedules in additional variables.  For the
case of a simple peptide we have demonstrated that our new technique offers
an easy way to improve the performance of  simulated annealing.

\vspace{0.5cm}
\noindent
{\bf Acknowledgements}: \\
This simulations were performed on the computers at
 the Institute for Molecular Science (IMS), Okazaki,
Japan.\\

%%%%%%%%%%%%%%%%%%%%%%%%% references %%%%%%%%%%%%%%%%%%%

\newpage
\noindent
{\Large Table 1:}\\
Results for simulated annealing simulations
using Tsallis weights as defined in Ref.~\cite{penna}. For each value of $q$ 20 
independent  runs of 50,000 sweeps were performed. The start temperature was
$T_{ST}=1000~K$. $q=1$ indicates  results from regular simulated annealing
which were taken from previous work in Ref.~\cite{ho94_3}. $n_G$ is the number 
of runs where a ground state configuration was found. $<E_{Low}>$ is the average
over 20 runs of the lowest energies obtained in each run. The standard 
deviation of this quantity is given in parentheses.

\begin{center}
\begin{tabular}{||c|c|c|c|c||}\hline \hline
       & \multicolumn{2}{c|}{ $T_F =50~K$ } &\multicolumn{2}{c||}{ $T_F = 1~K$ }
       \\ \hline
$q$    & $n_G$   & $<E_{Low}>$ & $n_G$& $<E_{Low}>$    \\ \hline
 2.00  &   0/20  & -6.0 (0.9)  & 2/20 &  -9.2 (1.3) \\
 1.75  &   0/20  & -7.5 (1.0)  & 1/20 &  -9.5 (1.2) \\
 1.50  &   3/20  & -9.6 (1.1)  & 1/20 &  -8.7 (1.3) \\
 1.25  &   5/20  & -9.5 (1.2)  & 4/20 &  -9.7 (1.4) \\
% 1.05  &   9/20  & -10.6 (1.3) & 9/20 &  -10.1 (1.8) \\
 1.00  &   6/20  & -10.0 (1.3) & 8/20 &  -10.0 (2.2)\\
 0.75  &   6/20  & -10.2 (1.3) & 7/20 &  -9.9 (1.9)\\ 
 0.50  &   5/20  & -10.2 (1.3) & 6/20 &  -9.9 (1.6)\\ 
 0.25  &   5/20  & -10.2 (1.3) & 8/20 & -10.0 (1.7)\\ 
 0.00  &   8/20  & -10.4 (1.3) & 2/20 & -9.0 (1.8) \\
 -0.50 &   5/20  & -9.3 (1.8)  & 5/20 & -9.1 (1.9) \\
 -1.00 &   4/20  & -9.5 (1.8)  & 5/20 & -9.6 (1.9) \\
 -2.00 &   1/20  & -8.6 (1.8)  & 1/20 & -8.1 (1.9) \\ \hline \hline
\end{tabular}
\end{center}

\newpage
\noindent
{\Large Table 2:}\\
Results for simulated annealing simulations
using Tsallis weights as defined in Ref.~\cite{straub}. For each annealing 
schedule characterized by the choice of start temperature $T_{ST}$ and final
temperature $T_F$  20 independent  runs of 50,000 sweeps were performed.  The
results are compared with that of regular simulated annealing runs. 
 $n_G$ is the number
of runs where a ground state configuration was found. $<E_{Low}>$ is the average
over 20 runs of the lowest energies obtained in each run. The standard
deviation of this quantity is given in parentheses.

\begin{center}
\begin{tabular}{||c|c|c|c|c|c||}\hline \hline
$T_{ST}/K$ &$T_{FI}/K$ &\multicolumn{2}{c|}{Regular Simulated Annealing}
  &\multicolumn{2}{c||}{Simulated Annealing Version of Ref.~5}\\ \hline
 &      & $n_G$ & $<E_{Low}>$      & $n_G$ & $<E_{Low}>$ \\ \hline
1000 & 50 & 6 & -10.0 (1.3)& 0 & -7.9 (1.8) \\ 
1000 & 1 & 8 & -10.0 (2.2)& 0 & -7.9 (1.3) \\
 500 & 50 & 8 & -10.5 (1.3)& 3 & -8.3 (1.8) \\
 500 & 1 & 2 & -9.3 (1.3) & 1 & -8.1 (1.6) \\ 
%400 &50 & 10 & -10.7 (1.4)& 2 & -8.2 (1.8) \\
% 400 & 1 & 2 & -9.3 (1.6) & 5 & -9.0 (2.0) \\ 
 300 & 50 & 1 & -9.8(1.2)  & 1 & -8.3 (1.4) \\
 300 & 1 & 3 & -9.6(1.4)  & 5 & -8.9 (2.0) \\ 
%  50 &50 & 0 & -6.6(1.4)  & 1 & -8.1 (1.3) \\
%   1 & 1  & 0 & -5.6(1.6)  & 4 & -8.9 (2.0) \\ 
\hline \hline
\end{tabular}
\end{center}

\newpage
\noindent
{\Large Table 3:}\\
Results for simulated annealing simulations
using Tsallis weights as defined in Eq.~\ref{wfinal}. For each annealing 
schedule characterized by the choice of start temperature $T_{ST}$ and  final
temperature $T_F$,
 20 independent  runs of 50,000 sweeps were performed.  The
results are compared with that of regular simulated annealing  runs. 
 $n_G$ is the number
of runs where a ground state configuration was found. $<E_{Low}>$ is the average
over 20 runs of the lowest energies obtained in each run. The standard
deviation of this quantity is given in parentheses.

\begin{center}
%\begin{tabular}{||c|c|c|c|c|c|c|c|c|c||}\hline \hline
% $T_{ST}/K$& $T_{FI}/K$   & \multicolumn{2}{c|}{RSA}
%   & \multicolumn{2}{c|}{ $q=1+1/19$} 
%   & \multicolumn{4}{c|}{  $1 \le q(T) \le 1+1/19$}\\ 
%   & \multicolumn{2}{c|}{ } 
%   & \multicolumn{2}{c|}{constant}
%   & \multicolumn{2}{c|}{exp.~increasing $q$} 
%& \multicolumn{2}{c|}{linear increasing  $q$}\\
%    \hline
% &  & $n_G$ &$ <E_{Low}> $ &$n_G$ & $<E_{Low}>$ &$n_G$ & $<E_{Low}>$ 
%                                             & $n_G$ & $<E_{Low}>$ \\ \hline
%1000 & 50   &6& -10.0 (1.3) &7&-10.7 (0.9)  & 7&-10.5 (0.9)  & 6&-10.5 (0.9)\\
%1000 & 1    &8& -10.0 (2.2) &7&-10.7 (1.3)  & 5&-9.9 (1.3)   & 2&-9.3 (1.3)\\ 
%500 &50     &8& -10.5 (1.3) &12&-11.1 (0.9) &14&-11.1 (0.9)  &11&-11.0 (0.9)\\
%500 & 1     &2& -9.3 (1.3)  &11&-10.9 (1.3) & 8&-9.6 (1.8)   & 5&-9.7 (1.9)\\
%300 &50     &5& -10.1 (1.3)  &13&-11.0 (0.9) &4&-9.7 (1.3)    &9&-11.0 (0.9)\\
%300 & 1     &3& -9.6 (1.4)  &11&-11.0 (1.1) & 3&-9.0 (1.7)   & 3&-9.4 (1.9)\\
\begin{tabular}{||c|c|c|c|c|c||}\hline \hline
$T_{ST}/K$ &$T_{FI}/K$ &\multicolumn{2}{c|}{Regular Simulated Annealing}
  &\multicolumn{2}{c||}{New Simulated Annealing Version}\\ \hline
 &      & $n_G$ & $<E_{Low}>$      & $n_G$ & $<E_{Low}>$ \\ \hline
 1000 & 50   &6& -10.0 (1.3) &7&-10.7 (0.9) \\
 1000 & 1    &8& -10.0 (2.2) &7&-10.7 (1.3) \\ 
 500 &50     &8& -10.5 (1.3) &12&-11.1 (0.9))\\
 500 & 1     &2& -9.3 (1.3)  &11&-10.9 (1.3))\\
 300 &50     &5& -10.1 (1.3)  &13&-11.0 (0.9)\\
 300 & 1     &3& -9.6 (1.4)  &11&-11.0 (1.1) \\
 \hline \hline
\end{tabular}
\end{center}

\end{document}